\documentclass[aps,prl,twocolumn,superscriptaddress,showpacs]{revtex4}
\def\be{\begin{equation}}
\def\ee{\end{equation}}
\def\bea{\begin{eqnarray}}
\def\eea{\end{eqnarray}}
\def\bi{\begin{itemize}}
\def\ei{\end{itemize}}

\usepackage{graphicx}
\usepackage{amsmath}
\usepackage{amssymb}

\begin{document}

\title{ Excitation energy after a smooth quench in a Luttinger liquid }

\author{Jacek Dziarmaga}
\author{Marek Tylutki}
\affiliation{
             Institute of Physics 
             and 
             Center for Complex Systems Research, 
             Jagiellonian University,
             Reymonta 4, 
             30-059 Krak\'ow, 
             Poland
}
\date{November 18, 2011}

\begin{abstract}
Low energy physics of quasi-one-dimensional ultracold atomic gases is often described by a gapless Luttinger liquid (LL). It is nowadays routine to manipulate these systems by changing their parameters in time but, no matter how slow the manipulation is, it must excite a gapless system. We study a smooth change of parameters of the LL (a smooth ``quench'') with a variable quench time and find that the excitation energy decays with an inverse power of the quench time. This universal exponent is $-2$ at zero temperature, and $-1$ for slow enough quenches at finite temperature. The smooth quench does not excite beyond the range of validity of the low energy LL description. 
\end{abstract}
\maketitle

\section{Introduction}
\begin{figure}[h!]
\includegraphics[width=0.8\columnwidth]{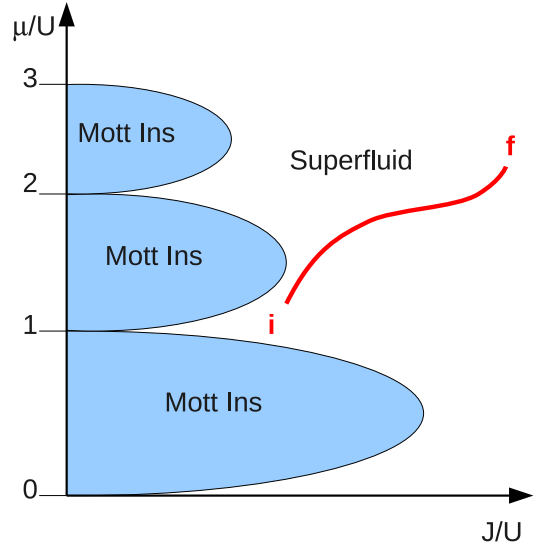}
\caption{{\it (color online).} The phase diagram of a Bose-Hubbard model (\ref{BH}). The superfluid regime, including the Mott insulator-superfluid boundary, has description in terms of the Luttinger liquid and any trajectory in the parameter space of the BH model maps to a trajectory in the $c-K$ plane of parameters of the Luttinger liquid. 
}
\label{fig.scheme}
\end{figure}

Quantum physics in one dimension (1D) is much different from that in higher dimensions. Many 1D quantum systems, both fermionic and bosonic, have low energy behaviour of a quantum Luttinger liquid (LL)~\cite{giamarchiBook}. No matter how complex is the underlying microscopic Hamiltonian, the effective quadratic LL Hamiltonian has only two parameters $c$ and $K$, where $c$ is a speed of its gapless excitations. In particular, LL description also applies to quasi-1D ultracold atom gases. For instance, it applies to a free Bose gas~\cite{monienBoseGas} and to bosons in a field of an optical lattice~\cite{monien}. In the case of the free Bose gas, the LL parameters can be related to those of the gas in the following way~\cite{monienBoseGas}: 
$
K = \sqrt{\frac{\kappa m}{\rho^3}},~ c = \sqrt{\frac{\kappa}{\rho m}}~, 
$
where $m$ is the mass of a boson, $\rho$ density, $\kappa$ compressibility, and $\hbar = 1$. In particular, LL also arises in the system of bosons interacting via contact interactions, where~\cite{LLquench,review}:
$
K = 1 + {4 \over \gamma},~ c = v_F (1 - {4 \over \gamma})~.
$
Here $\gamma=mg/\hbar^2\rho$ is interaction strength, with $g$ being the contact interaction's strength, and $\rho$ is linear density of particles. For $\gamma\to\infty$ we obtain 
a Tonks-Girardeau gas~\cite{Lieb} that was realized in the experiment of Kinoshita et al.~\cite{kinoshita}. Of special interest is a Bose gas in an optical lattice described by a Bose-Hubbard (BH) model
\be
H_{BH} = -J \sum_i (a^\dag_{i+1} a_i + {\rm H.c.}) + \frac{U}{2} \sum_i a^\dag_i a^\dag_i a_i a_i~.
\label{BH}
\ee
Its experimental realisation has been achieved e.g. in Refs.~\cite{greinerBloch,demarco}. When the ratio $J/U$ in~(\ref{BH}) is being varied, this model exhibits a quantum phase transition between the Mott insulator phase and the superfluid phase. For integer density we have Mott insulator pools in the $\mu-J$ phase diagram surrounded by the superfluid phase, see Figure \ref{fig.scheme}. The phase transition is of a commensurate-to-incommensurate type apart from situations where the integer density is kept fixed and the system undergoes a Berezinskii-Kosterlitz-Thouless (BKT) transition. These phase transitions can be described in terms of LL with $K=1$ for commensurate - incommensurate transition and $K=\frac12$ for BKT transition~\cite{fisher,monien}. The zero-gap superfluid phase can be also mapped to a LL Hamiltonian, and any transitions therein, driven by changing the ratio $J/U$, map to changes in the parameters $c$ and $K$ of the LL model. 

The ultracold gases are well isolated from their environment and it is easy to manipulate their Hamiltonian in time. Therefore, they can simulate dynamics of isolated quantum many-body systems driven by a time-dependent Hamiltonian, see Ref.~\cite{review} for reviews. In a typical experiment dedicated to the dynamics, a parameter in a Hamiltonian is suddenly switched between an initial and final value. However, even in a generic experiment, not dedicated the dynamics, a system is often manipulated by continuously turning its parameter from an initial to a desired final value in a ``quench'' time $\tau_Q$ (here we use the word ``quench'' too for historical reasons~\cite{KZ}). The idea is to prepare the system in a simple ground state of the initial Hamiltonian and then drive it adiabatically to the desired ground state of the final Hamiltonian. Unfortunately, this ideal adiabatic quantum state preparation must fail when the system is gapless either at an isolated quantum critical point or in a gapless phase \cite{review,nonadiab}. This failure can be quantified by e.g. excitation energy which is argued to decay with a power of the quench time $\tau_Q$~\cite{review,nonadiab}, often in accordance with the Kibble-Zurek (KZ) mechanism~\cite{KZ}. An example of this scaling was measured recently in a smooth quench from the Mott insulator to superfluid~\cite{demarco} and in the Dicke quantum phase transition~\cite{baumann}. 

Most theoretical work has been devoted to quenches across a quantum critical point~\cite{review}, where the scaling is determined by its critical exponents~\cite{KZ}. Recently, there has also been research on sudden interaction quenches in LL~\cite{LLquench,LLramp}. Since most experiments begin from Bose-Einstein condensation, much of the (supposedly adiabatic) manipulation is done in the gapless superfluid phase of a Bose-condensed system. With this motivation in mind, in the following we consider a smooth quench of parameters in a LL which is a universal low energy effective model for gapless (quasi-)1D quantum systems.

{\bf Luttinger liquid (LL).---} LL Hamiltonian is  
\be
H~=~
\frac12 c
\int_0^L dx~
\left[
K~\Pi^2~+~
K^{-1}~\left(\partial_x\Phi\right)^2
\right]~,
\label{HLL}
\ee
where $\Pi$ and $\Phi$ obey bosonic commutation relations,
\bea
\Pi(x) &=&
\sum_{k\neq0}
\left(\frac{|k|}{2L}\right)^{1/2}
\frac{k}{|k|}
e^{-ikx}
\left(a_k^\dag-a_{-k}\right)~,\\
\Phi(x) &=&
-i
\sum_{k\neq0}
\left(\frac{|k|}{2L}\right)^{1/2}
\frac{1}{k}
e^{-ikx}
\left(a_k^\dag+a_{-k}\right)~,
\eea
and $a_k$ and $a^\dag_k$ are bosonic annihilation and creation operators. A quench of one of the parameters in, say, the BH model (\ref{BH}) along a path in its superfluid phase, see Fig.~\ref{fig.scheme}, maps to a quench in the LL model (\ref{HLL}) along a path in the $c-K$ plane of its two parameters. This map is accurate provided the quench does not excite high energy states beyond the low-energy LL model.  

In terms of $a_k$ and $a_k^\dag$ the LL Hamiltonian (\ref{HLL}) reads
\bea
&&
H~=~
\sum_{k\neq0}~
c|k|~\times
\nonumber\\
&&
\left[
\left(\frac{K+K^{-1}}{2}\right) 
a_k^\dag a_k -
\left(\frac{K-K^{-1}}{2}\right)
\frac{a_ka_{-k}+h.c.}{2}
\right]~.
\label{HLLaadag}
\eea
For time-dependent $K(t)$ and $c(t)$ we make a Bogoliubov transformation
\be
a_k~=~u_k(t)~\gamma_k~+~v_{-k}(t)^*~\gamma_{-k}^\dag~
\label{Bog}
\ee
and assume that the state of the system is a Bogoliubov vacuum for $\gamma_k$'s. In this Heisenberg picture we have $i\frac{da_k}{dt}=\left[a_k,H_{\rm eff}\right]$.
Substituting Eq. (\ref{Bog}) and using $d\gamma_k/dt=0$ we obtain Bogoliubov-de Gennes equations
\bea
&&
i\frac{d}{dt}
\left(
\begin{array}{c}
u_k \\
v_k
\end{array}
\right)~=~
~c|k|~{\cal L}(K)~
\left(
\begin{array}{c}
u_k \\
v_k
\end{array}
\right)~,
\label{BdGt}\\
&&
{\cal L}(K)~=~
\frac12
\left(
\begin{array}{cc}
K+K^{-1} & K^{-1}-K \\ 
K-K^{-1} & -K-K^{-1}  
\end{array}
\right)~.
\eea
Instantaneous eigenmodes of ${\cal L}(K)$ with positive norm, $|u_k|^2-|v_k|^2=1$, 
\be
\left(u_k,v_k\right)~=~
\left(\frac{K+1}{2\sqrt{K}},\frac{K-1}{2\sqrt{K}}\right)~\equiv~
\left(U,V\right) 
\label{UV}
\ee
have positive instantaneous frequency $c|k|$. At the same time $(V,U)$ is an eigenmode of ${\cal L}(K)$
with negative norm, $|u_k|^2-|v_k|^2=-1$, and negative frequency $-c|k|$.

{\bf Quench at zero temperature.---}
We drive the Hamiltonian (\ref{HLLaadag}) by time-dependent $K(t/\tau_Q)$ and $c(t/\tau_Q)$. In the adiabatic basis (\ref{UV}) we have
\bea
\left(
\begin{array}{c}
u_k \\
v_k
\end{array}
\right)=
a_k
\left(
\begin{array}{c}
U \\
V
\end{array}
\right)
e^{-i|k|l(t)}+
b_k
\left(
\begin{array}{c}
V \\
U
\end{array}
\right)
e^{i|k|l(t)}~,
\label{Ansatz}
\eea
where $l(t)=\int^tdt'c(t')$, and Eq. (\ref{BdGt}) becomes
\bea
\frac{d}{ds} a_k &=& 
-b_k~e^{+2ic_f\tau_Q|k|s}~ \frac{d}{ds} \log K^{1/2}~,
\nonumber\\ 
\frac{d}{ds} b_k &=&
-a_k~e^{-2ic_f\tau_Q|k|s}~ \frac{d}{ds} \log K^{1/2}~.
\label{BdGab}
\eea
Here $s=\int^t\frac{dt'}{c_f\tau_Q}c\left(t'/\tau_Q\right)$ is dimensionless time-like variable. The amplitudes $a_k(s),b_k(s)$ satisfy $|a_k|^2-|b_k|^2=1$ and initial conditions $a_k(-\infty)=1,b_k(-\infty)=0$.

Average number of quasiparticles of momentum $k$ excited in the final state is $n_k=|b_k(\infty)|^2$. It depends on $k$ only through the product $c_f\tau_Q|k|$ defining a length scale
\be 
\xi~=~c_f\tau_Q~
\ee
which is the shortest wavelength of excited phonons. When $n_k$ decays with $|k|$ {\it sufficiently fast}, then average linear density of excited quasiparticles scales with $\tau_Q$ like
\be
n_{\rm ex}~=~\int_{-\Lambda}^{\Lambda}\frac{dk}{2\pi}~n_k~\sim~\tau_Q^{-1}~,
\label{nexgeneral}
\ee
while the more directly measurable excitation energy density scales like
\be
\varepsilon=\int_{-\Lambda}^{\Lambda}\frac{dk}{2\pi}~c_f|k|~n_k~\sim~\tau_Q^{-2}~ 
\label{energy}
\ee
provided that $\xi^{-1}\ll\Lambda$. These are universal scalings for quenches that do not excite beyond the range of validity of the LL model limited by the UV cut-off $\Lambda$.

\section{Adiabatic approximation}
\begin{figure}[h!]
\includegraphics[width=0.99\columnwidth,clip=true,angle=0]{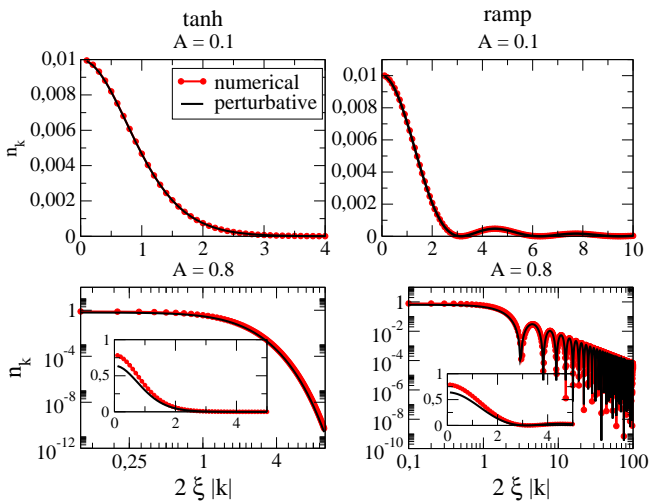}
\caption{{\it (color online).} 
The figure displays momentum dependence of excitation density $n_k$ as a function of $2\xi|k|$ for the tanh quench (left panels) in Eq. (\ref{tanh}) and the linear ramp (right panels) in Eq. (\ref{Ramp}). The upper row shows results for small quenches with $A = 0.1$ when the adiabatic approximation~(\ref{nkapprox}) agrees with numerical evaluation of equations~(\ref{BdGab}). For a greater $A=0.8$ (lower row) it remains valid for $2\xi|k|$ large enough, but not for small $2\xi|k|$ where it underestimates $n_k$ that tends to (\ref{sinh}) for $2\xi|k|\to0$ instead of $\log^2\sqrt{K_f/K_i}$ from a perturbative Eq.~(\ref{nkapprox}). The insets show corresponding plots in a linear scale.
}
\label{fig.momentum}
\end{figure}

In the most rapid limit of a sudden quench, considered in Ref.~\cite{LLquench}, we have
\be 
\lim_{c_f\tau_Q|k|\to0}n_k~=~
\sinh^2\left(\log\sqrt{ K_{f}/K_{i} }\right)~
\label{sinh}
\ee
which is small when the relative change of $K$ is small. Thus we can try an adiabatic approximation where $|b_k|^2\ll1$ and $a_k\approx1$. Solving Eqs. (\ref{BdGab}) perturbatively to leading order in $b_k$ yields a Fourier transform
\be
n_k~=~
\left|
\int_{-\infty}^{\infty} ds~
e^{-2i\xi|k|s}~
\frac{d\log K^{1/2}}{ds}
\right|^2~.
\label{nkapprox}
\ee
This $n_k$ is small and the adiabatic approximation is self-consistent when relative changes of $K$ during a quench are small. Moreover, even when they are large Eq. (\ref{nkapprox}) is still accurate for large enough $\xi|k|$ where $n_k$ is small, see the following examples and Figure \ref{fig.momentum}.

Here we consider three examples of quenches around a $K_0$: a smooth tanh quench 
\be 
\log K^{(1)}(s)/K_0~=~A~\tanh(s)~,
\label{tanh}
\ee
a linear ramp similar as in Ref.~\cite{LLramp}
\be 
\log K^{(2)}(s)/K_0~=~
A
\left\{
\begin{array}{cl}
-1 &, {\rm ~when~} s<-1,\\
s  &, {\rm ~when~} -1\leq s\leq 1,\\
1  &, {\rm ~when~} s>1,
\end{array}
\right.
\label{Ramp}
\ee
and a smooth shake beginning and ending at $K_0$
\be 
\log K^{(3)}(s)/K_0~=~\frac{A}{\cosh(s)}~.
\label{shake}
\ee
Corresponding power spectra (\ref{nkapprox}) are
\bea
n_k^{(1)}&=&
A^2
\left[
\frac{\pi\xi|k|}{\sinh\left(\pi\xi|k|\right)}
\right]^2~,
\nonumber
\\
n_k^{(2)}&=&
A^2
\left[
\frac{\sin\left(2\xi|k|\right)}{2\xi|k|}
\right]^2~,
\nonumber
\\
n_k^{(3)}&=&
A^2
\left[
\frac{\pi\xi|k|}{\cosh\left(\pi\xi|k|\right)}
\right]^2~.
\label{powersp}
\eea
They are exponentially localized on a scale $k\propto\xi^{-1}$ except the linear ramp (2) whose discontinuous time derivative results in a fat high energy tail $n_k^{(2)}\propto|k|^{-2}$. The linear density of excitations (\ref{nexgeneral}) reads
\bea 
n_{\rm ex}^{(1)}~=~ 
\frac{A^2}{6\xi},~
n_{\rm ex}^{(2)}~=~
\frac{A^2}{4\xi},~
n_{\rm ex}^{(3)}~=~
\frac{A^2}{12\xi},
\label{nex}
\eea 
when $\xi\Lambda\gg1$. All three densities scale like $\tau_Q^{-1}$ in agreement with the nonperturbative Eq. (\ref{nexgeneral}). 

The excitation energy density (\ref{energy}) reads
\bea 
\varepsilon^{(1)}&=&
\frac{3\zeta(3)}{2\pi^3}~\frac{A^2}{\xi^2}~,~~
\varepsilon^{(3)}~=~
\frac{9\zeta(3)}{8\pi^3}~\frac{A^2}{\xi^2}~,
\label{energyTanh}
\\
\varepsilon^{(2)}&=&
\frac{1}{8\pi}~\frac{A^2}{\xi^2}\log\left(2\xi\Lambda\right)~,
\label{energyRamp}
\eea
when $\xi\Lambda\gg1$. Here $\zeta$ is the Riemann zeta function. The energy scales like $\tau_Q^{-2}$, in agreement with the nonperturbative Eq. (\ref{energy}), except case (2) when it logarithmically diverges with the cut-off. The discontinuous time derivative of the linear ramp (\ref{Ramp}) is probing non-universal high energy excitations beyond the range of validity of the effective low-energy LL (\ref{HLL}). For the linear ramp the LL is not a self-consistent approximation and the scaling of the excitation energy with the ramp time $\tau_Q$, if any, can be different from the universal $\tau_Q^{-2}$. A non-universal exponent was observed in numerical simulations of linear ramps in the BH model~\cite{LLramp}.

\section{Correlations}
The excitation does not change the quasi-long-range character of correlation functions. For example, after a quench is completed a correlation function for density 
fluctuations smeared over distances longer than interparticle distance $\delta\rho(x)=\frac{1}{\pi}\partial_x\Phi(x)$ reads
\bea
&&
C(R)=
\langle \delta\rho(x+R) \delta\rho(x) \rangle~=~
\nonumber\\
&&
K_f
\int_{-\Lambda}^\Lambda\frac{|k|dk}{4\pi^3}
e^{ikR}
\left[1+2n_k+2\sqrt{n_k}\cos\varphi_k(t)\right],
\label{CR}
\eea
where $\varphi_k(t)=2c_f t|k|+\varphi^0(\xi|k|)$. In the ground state, when $n_k=0$, its tail decays like $C_f(R)=-\frac{K_f}{2\pi^3R^2}$ for $R\gg1/\Lambda$. The excitations $n_k>0$ add a correction 
\be 
\frac{K_f}{\xi^2}
\left[
F\left(\frac{R}{\xi}\right)+
G\left(\frac{2c_f t-R}{\xi}\right)+
G\left(\frac{2c_f t+R}{\xi}\right)
\right].
\label{dCR}
\ee
Here $F(z)$ and $G(z)$ are real functions originating from the $n_k$ and $\sqrt{n_k}$ terms in Eq.~(\ref{CR}) respectively . 

There is analogy to the quasiparticle horizon effect~\cite{Horizon}. The $G$-terms in (\ref{dCR}) describe shock waves that originate from correlated pairs of quasiparticles with momenta $\pm k$ excited during a quench whose separation grows like $2c_ft$. Their width $\simeq\xi$ is the shortest length on which quasiparticles excited in time $\tau_Q$ can be localized in space. The $F$-term in (\ref{dCR}) is an additive correction to $C_f(R)$ that remains inside the quasiparticle horizon, $R\ll 2c_ft$, after the shock waves go away. When $n_0>0$ in (\ref{sinh}), then a tail of the remaining correlator is $C(R)=(1+2n_0)C_f(R)$ for $R\gg\xi$. The $R^{-2}$ tail of the ground state correlator is amplified by a factor $(1+2n_0)=\frac12(K_f/K_i+K_i/K_f)$. When $n_0=0$ in (\ref{sinh}), as for e.g. the shake (\ref{shake},\ref{powersp}), then the $F$ term decays faster than $R^{-2}$ and the tail of $C(R)$ is the same as in the ground state: $C(R)=C_f(R)$ for $\xi\ll R \ll 2c_ft$.  

Despite all its interesting physics, the additive correction (\ref{dCR}) does not cut the algebraic $R^{-2}$-tail of the correlator in the quasi-long-range ordered ground state. This contrasts with the KZ mechanism where correlations after a quench from a disordered to an ordered phase decay exponentially~\cite{review}.

\section{Quench at finite temperature}
Since real experiments are done at finite temperature, we generalize to finite $T$, where the excitation spectrum reads
\be 
n_k(T)~=~n_k^{\rm BE}~+~n_k~\left(1+2n_k^{\rm BE}\right)~.
\label{nkT}
\ee
Here $n_k=|b_k(\infty)|^2$ is the excitation spectrum at $T=0$, and
$n_k^{\rm BE}=\left[\exp(c_i|k|/T)-1\right]^{-1}$ is the initial thermal distribution which can be also reinterpreted as a final thermal distribution 
$n_k^{\rm BE}=\left[\exp(c_f|k|/T_f)-1\right]^{-1}$ with 
\be 
T_f~=~\frac{c_f}{c_i}T~.
\ee
This temperature rescaling is the only effect in the adiabatic limit $\tau_Q\to\infty$ when $n_k\to 0$.

Here we are more interested in the non-adiabatic excitation above this thermal background: 
\be 
n_k(T)-n_k^{\rm BE}~=~n_k~\left(1+2n_k^{\rm BE}\right)~.
\ee
When compared with the bare $n_k$ at $T=0$, it is amplified by a Bose enhancement factor $1+2n_k^{\rm BE}$ making the excitation more significant than at $T=0$. For relatively fast quenches, with $\tau_QT\ll1$, the distribution $n_k$ extends up to $|k|\simeq\xi^{-1}$ where the enhancement $1+ 2 n_k^{\rm BE}\approx1$. Consequently, the non-adiabatic excitation energy density $\varepsilon(T)$ is roughly same as at $T=0$. In contrast, for relatively slow quenches with
\be 
\tau_Q~T~\gg~1~,
\ee  
the $k\to0$ singularity of the Bose enhancement factor has more qualitative consequences. In this regime the non-adiabatic excitation energy density is
\bea
\varepsilon(T) 
&=&
\int_{-\Lambda}^{\Lambda}\frac{dk}{2\pi}~
c_f|k|~
n_k\left(1+2n_k^{\rm BE}\right)
\nonumber\\
&\approx&
\int_{-\Lambda}^{\Lambda}\frac{dk}{2\pi}~
c_f|k|~
n_k~
\frac{2T_f}{c_f|k|}~=~
2T_f~n_{\rm ex}~,
\label{energyT}
\eea
where $n_{\rm ex}$ is the density of excited quasiparticles at $T=0$ in (\ref{nexgeneral}). The energy is not only larger than its $T=0$ counterpart (\ref{energy}), but its decay with $\tau_Q$ is also less steep:
\be 
\varepsilon(T) ~\sim~ \tau_Q^{-1}~
\ee
from (\ref{nexgeneral},\ref{energyT}) instead of the $\tau_Q^{-2}$ at $T=0$ in (\ref{energy}).

\section{Conclusion} 
We derived universal dynamical exponents for the scaling of excitation energy with a quench time when the quench is slow and smooth enough not to excite beyond the quadratic Luttinger liquid model. Due to the singularity of the Bose enhancement factor, the exponents are different at zero and finite temperature. In both regimes they can be used to test the low energy LL description of quasi-1D ultracold atomic systems.

{\it Acknowledgments.} Work supported by NCN grant DEC-2011/01/B/ST3/00512 (J.D.) 
and Polish Government project N202 124736 (M.T.).


\end{document}